\documentclass[aps,prb,twocolumn,showpacs,groupedaddress,amsmath,amssymb]{revtex4}

\usepackage{graphicx}

\begin{document}
\bibliographystyle{prsty}
\input epsf
\title{Effect of Increasing Disorder on the Critical Behavior of a Coulomb 
System}
\author{Michael H. Overlin, Lee A. Wong and Clare C. Yu}
\affiliation{
Department of Physics and Astronomy, University of California,
Irvine, Irvine, California 92697}
\date{\today}
\begin{abstract}
We have performed a Monte Carlo study of a classical three dimensional
Coulomb system in which we systematically increase the positional disorder.
We start from a completely ordered system and gradually
transition to a Coulomb glass.
The phase transition as a function of temperature is second order for 
all values of disorder.
We use finite size scaling to determine the transition temperature $T_C$ and 
the critical exponent $\nu$. We find that $T_C$ decreases and that $\nu$
increases with increasing disorder. We also observe changes in the specific
heat, the single particle density of states, and the staggered 
occupation as a function of disorder and temperature.

\end{abstract}

\maketitle

\section{Introduction}
Electrons with long range Coulomb interactions 
display a rich and complex behavior. 
In doped semiconductors and disordered metals, electrons are
in the presence of quenched disorder, and the competition between Coulomb
interactions and disorder produces a Coulomb glass which is
an amorphous insulator. 
A great deal of effort has been expended in studying various thermodynamic
properties of Coulomb glasses such as the specific heat
\cite{Mobius97,Mobius01}, the presence of a Coulomb glass phase
transition in which the electrons are frozen into
a highly disordered arrangement
\cite{Davies82,Davies84,Grunewald82,Grannan93,Vojta93b},
and the Coulomb gap \cite{Pollak70,Efros75,efrosbook}.
Coulomb interactions between localized
electrons result in the so--called Coulomb gap in the single
particle density of states that is centered at the Fermi energy.
Simulations have found a Coulomb gap in the density of states
\cite{Levin87,Grannan93,Li94,Mogilyanskii89,Vojta93,Sarvestani95}, and
experimental evidence for a Coulomb gap has been seen in tunneling
measurements \cite{Massey95,Massey96,Lee99,Sandow01}.

Many of the theoretical studies of Coulomb glasses
have been as a function of temperature.
In this paper we will study what happens as we vary
the amount of disorder as well as the temperature.
We will start with an ordered system and study the effect of gradually
introducing disorder into a three dimensional system of electrons with long
range Coulomb interactions. The system is discrete in the sense 
that the electrons
sit on half of the available sites. In the ordered case the sites form a cubic
lattice. The disorder is introduced in the positions of the sites and their
deviation from the positions in a cubic lattice. For all values of
disorder, the system undergoes a second
order phase transition as the temperature is lowered. We will 
study the effects on
the thermodynamics of this phase transition as a function of disorder.

Discretizing our Coulomb system means that it corresponds to an Ising
system with long range interactions. An site occupied with an
electron corresponds to spin-up and an empty site corresponds to spin-down.
This is a very general model, and as a result 
relevant work has been done in other fields motivated by somewhat different
physical systems. In particular there is the Ising model with long
range interactions. Also the ordered case is related to work that
has been done on ionic fluids near criticality. It is worth briefly
reviewing the work that has been done in those fields. 

In the case of translational 
invariance, ionic fluids near criticality have been a subject of 
both experimental and theoretical investigations 
\cite{Ciach02,Panagiotopoulos99,Dickman99,Brognara02}.
As in the case of electrons, this system is
somewhat simplified by discretizing the
system and only allowing the charges to sit on specified sites.
For ionic fluids this is known as the lattice restricted primitive model
(LRPM) \cite{Panagiotopoulos99,Dickman99,Brognara02} 
where there are equal numbers of positive
and negative ions with the same diameter sitting on lattice sites.
In the LRPM there is
no quenched disorder. There are positive sites, negative sites,
and neutral sites (empty sites) corresponding to an Ising spin--1
model with Coulomb interactions. The phase diagram in the density--temperature
plane has a second order transition line from a high temperature
paramagnetic phase to a low temperature antiferromagnetic phase
\cite{Panagiotopoulos99,Dickman99,Brognara02}. This transition
is in the Ising universality class with critical exponent $\nu=0.63$.
At even lower temperatures there is a first order phase transition in
which the system undergoes a phase separation into a high density
ordered phase and a low density disordered phase. If there are no
neutral sites (ionic density $\rho=1$), which corresponds to 
the antiferromagnetic spin-1/2 Ising model,
then there is just the second order transition from the high temperature
disordered phase to the low temperature ordered antiferromagnetic phase
in three dimensions.  For the purposes of this 
paper we are interested in this case where there are no neutral
sites. Every positively charged site has a positive ion or missing electron,
and every negatively charged site has a negative ion or an electron.
The fact that the ionic system has a second order phase 
transition to an ordered antiferroelectric arrangement of ions
\cite{Panagiotopoulos99,Dickman99,Brognara02} means that we 
expect the analogous transition to occur for the case where the 
electrons can sit on alternate lattice sites with no quenched disorder. 

Comparing the ordered and disordered extremes reveals similarities and
differences. Both systems undergo a phase transition when the temperature
is lowered. In the ordered case, the transition is to an ordered arrangement
of electrons occupying every other site
whereas in the disordered case the electrons are frozen into
the highly disordered arrangement of a Coulomb glass \cite{Grannan93}. 
Both systems at low temperatures
have a gap in their single particle density of states. 

As we mentioned earlier, systems with either positively charged sites
(missing electron) or negatively charged sites (electron present) 
can be mapped onto Ising spin-1/2 systems. A great deal of
work has been done on Ising models with long range interactions.
The ferromagnetic or attractive
Ising model with power law interactions that fall off as $1/r^{\eta}$
without quenched disorder has been studied \cite{Fisher72,Luijten02}
as a function of the
dimension $d$ and the exponent $\eta$ for $\eta\geq d$.
However, in this paper we will focus
on interacting electrons and so we are interested in the 
antiferromagnetic Ising model.

The presence of quenched disorder results in an Ising spin glass. 
There has been a substantial amount of numerical effort to understand the 
energy of domain walls at $T=0$. 
\cite{Bray84,McMillan84a,McMillan84b,Rieger96,Hartmann01,Carter02,Katzgraber03a,Katzgraber03b} 
The energy of the domain wall goes as
$L^{\theta}$ where $L$ is the system size and the exponent
$\theta$ is positive for systems with nonzero transition temperatures.
Work on the Ising spin glass with power law interactions has been summarized
in a couple of papers \cite{Fisher88,Katzgraber03a}. The
system has a rich phase diagram in
the $d-\eta$ plane which can be found in Ref.~\cite{Katzgraber03a}, where
$d$ is the dimension and $\eta$ is the exponent of the power law
interaction $1/r^{\eta}$. The smaller $\eta$ is, the longer the
range of the interaction. If the range is long enough or if
the dimension is large enough, then there is a second order
phase transition with a transition temperature $T_C > 0$.
The critical exponents are different in the long range and 
short range regimes. The exponent $\theta$
depends continuously on $\eta$ in the long-range region 
($\theta=d-\eta$), and is
independent of $\eta$ in the short-range regime \cite{Katzgraber03a}.
This indicates that the critical exponents also depend
continuously on $\eta$ in the long-range region, and
are independent of $\eta$ in the short-range regime \cite{Katzgraber03a}.
Katzgraber and Young have done Monte Carlo simulations 
of an Ising spin glass in
one dimension with long range interactions 
\cite{Katzgraber03a,Katzgraber03b}. They chose a value for
$\eta$ where the system has a second order
spin glass transition, and they find that $\nu=10/3$.

In this paper we will be concerned with what happens to thermodynamic
quantities as we
systematically introduce disorder into a three dimensional system 
of electrons with long range Coulomb interactions. 
The disorder is introduced into the placement of sites where the
electrons can sit.
The paper is organized as follows. In section II we present the Hamiltonian
and describe our Monte Carlo simulation. In section III we present the
quantities that we measure. In section IV we present our results, and we give
our conclusions in section V.

\section{Calculation}
\subsection{Hamiltonian}
Let us start by considering the completely disordered case which is known as a Coulomb
glass.
The essential physics of the Coulomb glass is the presence of both disorder and long
range Coulomb interactions between electrons. The Hamiltonian often studied for the
Coulomb glass is \cite{Davies82,Baranovskii79}
\begin{equation}
H=\sum_{i}n_i \phi_i + \sum_{i>j}\frac{\left(n_i-K\right)\left(n_j-K\right)}{r_{ij}}
\end{equation}
where we set the charge $e=1$, $n_i=\pm 1$ is the number operator for site $i$, 
$\phi_i$ is the onsite energy, $r_{ij}=|\vec{r}_i-\vec{r}_j|$, and
$K$ is a compensating background charge making the whole system charge
neutral. Such a Hamiltonian describes a lightly doped semiconductor, in
which the impurity sites are far enough apart that the overlap between sites
can be neglected. In most of the early work on the Coulomb glass (e.g.,
Refs. \cite{Davies82,Grunewald82,Baranovskii79}), the sites are chosen
to form a periodic lattice, and the disorder is present in the form of
random onsite energies. For an ordered system, the onsite energy
$\phi_i$ is a constant. One could imagine gradually introducing disorder 
by allowing $\phi_i$ to be chosen from a distribution whose width gradually 
increases.

However, the presence of random onsite energies makes numerical analysis 
difficult,
since even in the high temperature state the average occupation of a site is
not zero. This makes the search for a phase transition difficult; there is no 
obvious order parameter which becomes nonzero at the transition. 
For our numerical analysis, it is more convenient to take the disorder to be
entirely in the location of the sites. This changes the symmetry
of the Hamiltonian from having onsite disorder to having 
disorder in the interaction between sites because the distance
between sites varies. For many quantities these two models give
similar results. For example studies of the specific heat in Coulomb glasses
have compared having disorder in the onsite energy to 
having a completely random displacement of sites
\cite{Diaz00,Mobius01}. They find that both
models produce qualitatively similar results with some
quantitative differences. However, the existence and nature of
the phase transitions is different in the two models 
\cite{Vojta94,Grannan94}.
In particular there is always a phase transition no matter how
wide the distribution of the site placement is, whereas
there is no phase transition if the width of 
the distribution of the onsite energy 
$\phi_i$ is larger than a critical value \cite{Mobius04}.
M{\"o}bius \cite{Mobius04} has argued that such a critical value 
must exist, even if it is vanishingly small, since there is a phase 
transition when there is no disorder \cite{Mobius03} while there is 
no clear evidence for a transition when
there is substantial onsite disorder. This implies that long 
range order is destroyed by both onsite disorder and thermal fluctuations.  

A number of previous simulations have used the 
form for the Hamiltonian with disorder in the placement of the sites
\cite{Xue88,Grannan93,Diaz00,Mobius01}.
In the case of half filling there is a particle--hole symmetry, and 
the phase transition is associated with the development of a nonzero 
Edwards-Anderson order parameter \cite{Grannan93}. We therefore rewrite
the Hamiltonian (taking $K=1/2$) to look like that of a spin glass,
\begin{equation}
H=\frac{1}{4}\sum_{i>j}\frac{S_i S_j}{r_{ij}}
\label{eq:Ham}
\end{equation}
$S_i=1$ ($-1$) will denote an occupied (unoccupied) site. 

We have simulated three dimensional systems of linear size $L=4$, 6, and 8. 
We place $N=L^3$ sites in the system. We have only considered
the case of half filling in order to take advantage of the
spin-flip symmetry. For the ordered case
the sites form a cubic lattice. In the ground state, every other site is 
occupied; the occupied sites form
a face centered cubic (FCC) lattice. We can gradually introduce disorder
by allowing the deviation of a site from its position in a cubic lattice 
to be chosen from a Gaussian distribution with a standard deviation of 
$\sigma$. This gives the radial distance from the cubic lattice site.
The angular coordinates of the site are chosen randomly using a uniform
distribution. The 
ordered case corresponds to $\sigma=0$. $\sigma=1$ corresponds to the very 
disordered case with a standard deviation equal to the cubic lattice constant $a$.
We also considered completely random arrangements of sites where the
$x$, $y$, and $z$ coordinates of each are chosen from a uniform distribution.
We call this the ``uniform random'' case. We find no qualitative
difference and only a slight quantitative difference between
the uniform random case and the $\sigma=1$ case in quantities such
as the single particle density of states, the specific heat versus
temperature, and the Binder's $g$. So we will not make much mention
of the uniform random case.

We use infinite periodic boundary conditions in which the simulation box is
infinitely replicated in all directions to form a lattice. As a result,
an electron on a given site interacts with other electrons and all their
images via the Coulomb interaction. To handle this, we use an Ewald summation 
technique \cite{DeLeeuw80} which replaces the Hamiltonian in Eq. (\ref{eq:Ham}) 
with the following effective interaction between sites:
\begin{equation}
H=\sum_{1\leq i < j\leq N}\frac{1}{L}q_i q_j\psi\left(\frac{\vec{r}_{ij}}{L}
\right)+\frac{\Lambda}{2L}\sum_{i=1}^{N}q_i^2 
\label{eq:Ewald}
\end{equation}
where $L$ is the linear size of the simulation box, $N$ is the number of sites,
the charge $q_i=S_i/2$,
and the function $\psi(\vec{r})$ is given by
\begin{eqnarray}
\psi(\vec{r})& = &\sum_n \frac{{\rm erfc}(\alpha|\vec{r}+\vec{n}|)}
{|\vec{r}+\vec{n}|} \nonumber\\
& + &\frac{1}{\pi}\sum_{n\neq 0}\frac{1}{|\vec{n}|^2}
\exp\left\{2\pi i\vec{n}\cdot\vec{r} - \frac{\pi^2 |\vec{n}|^2}{\alpha^2}\right\}
\end{eqnarray}
in which
\begin{equation}
{\rm erfc}(x)=1-\frac{2}{\sqrt{\pi}}\int_{0}^{x}e^{-t^2}dt
\label{eq:erfc}
\end{equation}
is the complementary error function and
\begin{equation}
\Lambda=\sum_{\vec{n}\neq 0}\left[\frac{{\rm erfc}(\alpha|\vec{n}|)}{|\vec{n}|}
+\frac{1}{\pi|\vec{n}|^2}e^{-\pi^2 n^2/\alpha^{2}}\right]-
\frac{2\alpha}{\sqrt{\pi}}
\label{eq:Lambda}
\end{equation}
Note that
\begin{equation}
\Lambda=\lim_{|r|\rightarrow 0}\left[\psi(\vec{r})-\frac{1}{|\vec{r}|}\right]
\end{equation}
The sum over $\vec{n}$ in Eq.~(\ref{eq:Lambda}) is a sum over all 
simple cubic lattice 
points with integer coordinates $\vec{n}=(l,m,n)$. These are the coordinates of the images
of the simulation box. The parameter $\alpha$ is a convergence factor that
is adjusted to maximize the rate of convergence of the sum. We have omitted
a positive term in the Hamiltonian (\ref{eq:Ewald}) that is proportional to 
the square of the net dipole moment of the configuration \cite{DeLeeuw80}. 
This omission is equivalent to the boundary condition in which the 
infinite sphere of our system and its images
is surrounded by a perfect conducting medium. We have done some runs with the
dipole term and find no qualitative difference and only a slight 
quantitative difference compared to the case with no dipole term.
So in this paper we will present the result of simulation runs which omit the
dipole term.

\subsection{Monte Carlo Simulation}
We have used a Monte Carlo heat bath algorithm. We keep a table of the 
potential energy at each site. Each electron is looked at sequentially
and moved to one of the available $N/2 + 1$ sites (its own site or one
of the available $N/2$ unoccupied sites), chosen with a Boltzmann
probability. If the site chosen is the electron's original location,
the potential energies are unchanged; if the electron hops to a new
site, we update all the potential energies. If the electron chooses
its initial site, which it does with high probability at low temperatures,
we do not have to recompute the potential energies. This speeds up
the simulation considerably, partially compensating for the much longer
equilibration times needed at low temperatures. Our longest run
(for $L=4$ at $T=0.01$ and $\sigma=0.5$) had $3\times 10^6$ Monte 
Carlo steps per electron.
Depending on the system size and temperature, the sample averages
involved between 5 and 190 disorder configurations.

\section{Measured Quantities}
\subsection{Binder's $g$ and equilibration criteria}
The Edwards--Anderson order parameter alluded to above quantifies
the extent to which spins or site occupations are frozen. It is defined as 
$q\equiv \left[\langle S_i\rangle^2\right]$; we will denote thermal
averages by $\langle\; ...\;\rangle$ and disorder averages by $[\; ...\; ]$.
Thermal averages sum over fluctuations in the positions of the electrons
weighted with the correct Boltzmann probability; disorder averages sum over
different arrangements of the sites. We can see from the definition of $q$ that
if the spins are frozen, then the average orientation of a spin will have a nonzero 
thermal average and $q$ will be finite. This is why $q$ can be thought of as
the order parameter of the phase transition.

We can generalize the Edwards--Anderson order parameter to a finite time
overlap in either of two ways \cite{Bhatt88}. The first way computes the 
overlap between two replicas:
\begin{equation}
q_r(t)=\frac{1}{N}\sum_i S_i^{(1)}(t) S_i^{(2)}(t)
\end{equation}
where the superscripts refer to different replicas. The two replicas
are identical in their disorder, i.e., the placement of the sites, but
differ in the initial positions of the electrons. The other way
uses the same replica at two different times
\begin{equation}
q_t(t,\tau)=\frac{1}{N}\sum_{i}S_i^{(1)}(t) S_i^{(1)}(t+\tau).
\end{equation}
If the time difference $\tau$ is sufficiently large that the 
electron configurations at $t$ and $t+\tau$
are essentially uncorrelated, $q_t(t,\tau)$ will give the same 
result as the replica overlap. 

We use the moments of the overlap to define  
Binder's $g$ which is a parameter that is related to the phase transition. 
First we define  $g(\tau)$ by \cite{Binder81,Bhatt88}
\begin{equation}
g(\tau)=\frac{1}{2}\left(3-\frac{\left[\langle q^4(\tau)\rangle\right]}
{\left[\langle q^2(\tau) \rangle\right]^2}\right)
\end{equation}
where
\begin{equation}
\langle q^n(\tau)\rangle=\frac{1}{\tau}\sum_{t=\tau}^{2\tau} q^n(t)
\end{equation}
We will use $g_t$ ($g_r$) to denote the result of using $q_t$
($q_r$). Binder's $g$ is given by
\begin{equation}
g=\lim_{\tau\rightarrow\infty} g(\tau)
\end{equation}
which we will approximate by
\begin{equation}
g\approx g(\tau)
\end{equation}
for some measurement time $\tau$ large enough that the configurations
are essentially uncorrelated so that $g_t$ and $g_r$ agree. We have used
this fact to monitor equilibration by simulating two replicas that have
the same placement of sites but different spin configurations.\cite{Bhatt88}
Typically $g_r(t)$ increases with time to the equilibrium value,
whereas $g_t(t)$ decreases to the equilibrium
value. The two methods agree when the system has reached equilibrium.
Our criterion for equilibration was that the values of $g_r$ and $g_t$
agreed to within 0.1. We only present results for systems which meet
this criterion. We should caution that this criterion can be met even
though the system may still be slowly aging. 

Binder's $g$ provides a way to monitor the phase transition.
At high temperatures, the distribution of $q$ tends to a Gaussian so that
$g\rightarrow 0$, whereas the
order parameter, and hence $g$, become nonzero as the temperature
approaches the phase transition temperature $T_C$. If we make
the assumption of one parameter scaling, then the only relevant length is 
the correlation length 
$\xi\sim \left(T-T_{C}\right)^{-\nu}$ where $\nu$ is the critical exponent
associated with $\xi$. So all lengths, including $L$, can be scaled by $\xi$.
Since $g$ is dimensionless, we expect that it should satisfy a scaling form
\cite{Binder81,Bhatt88}
\begin{equation}
g(L,T)=\hat{g}\left(L^{1/\nu}\left(T-T_C\right)\right).
\label{eq:scaling_g}
\end{equation}
Thus at the critical temperature, $g(L,T_C)$ should have the same
value independent of the system size $L$ (as long as $L$ is 
sufficiently large for finite size scaling to apply).\cite{Binder81,Bhatt88}
 
\subsection{Specific Heat}
There are two ways to calculate the specific heat $C_V(T)$. The first way 
uses the variance of the energy fluctuations:
\begin{equation}
C_V=\frac{1}{Nk_B T^2}\left[\langle E^2 \rangle - \langle E \rangle^2\right]
\label{eq:sphtVar}
\end{equation}
where $E$ is the average energy per electron, $N$ is the number of electrons, 
and $k_B$ is Boltzmann's constant. The other way to calculate the specific
heat is to take the derivative of the energy with respect to temperature. We
can approximate the derivative by a finite temperature difference
\begin{eqnarray}
C_V(T_i)&=&\left.\frac{\partial\left[\langle E\rangle\right]}{\partial T}\right|_{T_i}
\nonumber \\
&\approx& \frac{\left[\langle E(T_{i+1})\rangle\right]-
\left[\langle E(T_{i})\rangle\right]}{\left(T_{i+1}-T_{i}\right)}
\label{eq:sphtDer}
\end{eqnarray}
We found that the specific heat calculated in these two ways agreed quite 
well. Notice
however, that if the slope of $E$ versus $T$ is increasing as temperature increases, 
then the specific heat calculated by the finite difference method will underestimate
$C_V$ which is actually the slope of the tangent to the energy curve. 
Similarly, if the slope of $E$ versus $T$ is decreasing as the
temperature increases, the finite difference method will overestimate $C_V$.

\subsection{Staggered Occupation}
Since an unoccupied site on a cubic lattice corresponds to a down Ising spin
and an occupied site to an up Ising spin, the FCC crystalline phase corresponds
to a maximum in the magnitude of the staggered occupation $M_s$. 
The staggered occupation is defined by \cite{Vojta93b}
\begin{equation}
M_s=\frac{1}{N}\sum_i (-1)^{i+j+k} S_{i+j+k}
\label{eq:stagmag}
\end{equation}
where $i$, $j$, $k$ are the integer coordinates
of the sites in a cubic lattice in units of the lattice constant $a$.
So a site coordinate $(x,y,z)=(ia,ja,ka)$.
Since disorder is introduced through a distribution in the position of the
sites with respect to the cubic lattice sites, we can still use
eq. (\ref{eq:stagmag}) to calculate the staggered occupation in the presence of 
disorder by regarding $i$, $j$, and $k$ as coordinates of the center of the 
unit cell where the site is located. It is useful to plot the 
staggered occupation distribution
$P(M_s)$ versus $M_s$ in order to see the extent of the 
``crystalline'' order. In order to compare different system sizes, we normalize
the staggered occupation to range from $-1$ to $+1$, and the
area under the curve is normalized to 1.

\subsection{Single Particle Density of States}
We have calculated the single particle density of states $N(E)$ 
at various temperatures. $N(E)$ is the distribution of potential 
energies at single sites due to interactions with all the other 
sites. In other words, we can write the Hamiltonian in the form
of an Ising model:
\begin{eqnarray}
H&=&-\frac{1}{2}\sum_{ij}J_{ij}S_i S_j \nonumber\\
&=&\sum_{i}E_i S_i
\end{eqnarray}
where $E_i$ is the single site energy or ``local field'' and is given by 
\begin{equation}
E_{i}=-\frac{1}{2}\sum_{j}J_{ij}S_j
\end{equation}
$N(E)$ is the thermally averaged and disorder averaged distribution of $E_i$. 

\section{Results}
\subsection{Second order melting transition in the ordered case ($\sigma=0$)}
We consider the case where the translational degrees
of freedom are discrete and the electrons can only sit on designated sites.
This is equivalent to a long range Ising model in three dimensions. If the
sites are ordered and are lattice sites, it is
also equivalent to the lattice restricted primitive model (LRPM) in the
completely filled case (ionic density $\rho=1$) \cite{Brognara02}. We find
that discretization produces a second order phase transition regardless of
the amount of positional disorder. In this section we present evidence
that the ordered case with $\sigma=0$
undergoes a second order crystallization transition to an FCC lattice as
the temperature is lowered. This result is consistent with the
second order transition found by M{\"o}bius and R{\"o}{\ss}ler 
\cite{Mobius03} from numerical simulations of a half--filled
system on a cubic lattice with Coulomb interactions. It is also
consistent with the second order transition in the LRPM model for the fully 
occupied case ($\rho=1$) \cite{Brognara02,privCommCiach}.

\begin{figure}
\includegraphics[width=3in]{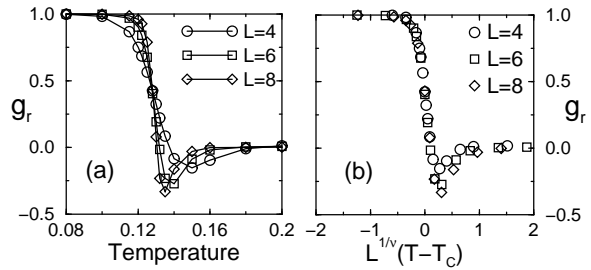}
\caption{(a) $g_r(L,T)$ vs. $T$ for $\sigma=0$. 
($g_t(L,T)$ vs. $T$ is virtually identical.)
The data for $L=4$ is averaged over 190 runs, $L=6$ is averaged over 67 runs,
and $L=8$ is averaged over 45 runs. The solid lines are guides to the eye.
(b) $g(L,T)$ for $\sigma=0$ scaled using 
$\hat{g}\left(L^{1/\nu}\left(T-T_C\right)\right)$ with $T_C=0.128\pm 0.001$ and
$\nu=0.55\pm 0.1$.
}
\label{fig:gr_sigma=0}
\end{figure}
In Figure \ref{fig:gr_sigma=0} we show $g$ versus the temperature $T$
at $L=4$, 6, and 8. The point where these curves cross yields a transition
temperature of $T_C=0.127$. Notice that above the transition $g$ dips down
and acquires negative values. This behavior has been seen in the case of
a 3--state ferromagnetic Potts model in three dimensions which undergoes a 
first order phase transition \cite{Vollmayr93}. However, in that case
the value of $g$ at the minimum scaled as $g(T_{\rm min}) \sim -L^d$, whereas
in our case $g(T_{\rm min})$ appears to saturate at large $L$. The negative
values of $g$ can result if the distribution $P(q)$ is nongaussian
with finite weight at $q\neq 0$ corresponding to long lived occupations of
some sites. A very simple delta function distribution that illustrates this is  
\begin{equation}
P(q)=\alpha_o\delta(q)+\left[\frac{1-\alpha_o}{2}\right]\delta(q-a_o)+
\left[\frac{1-\alpha_o}{2}\right]\delta(q+a_o)
\end{equation}
where $\alpha_o$ is a parameter with values between 0 and 1, 
and $a_o$ is a constant. For $2/3 < \alpha_o <1$,
this distribution yields $g<0$.

We initially thought that the transition might be first order.
One of the signatures of a first order melting transition is coexistence of the
liquid and crystalline phases at the melting temperature. We looked
for evidence of coexistence by examining the distribution $P(M_s)$ of
the staggered occupation. Coexistence would produce three peaks in $P(M_s)$ 
versus $M_s$: a central peak and two side peaks symmetrically placed 
with respect to $M_s=0$. The central 
peak corresponds to the high temperature liquid phase and the
side peaks correspond to the FCC crystalline phase. Furthermore, at
the transition temperature for a first
order transition, the three peaks would become narrower and higher with
increasing system size. On the other hand, if the system is cooled
through a second order transition, the high temperature central peak in $P(M_s)$
is replaced by two peaks symmetrically placed about $M_s=0$.
These peaks do not become sharper with increasing system size, but the
width of the distribution is expected to decrease with increasing system size 
as $L^{-\beta/\nu}$ where $\beta$ and $\nu$ are the critical exponents 
defined by $M_s\sim\left|T-T_C\right|^{\beta}$ and 
$\xi\sim\left(T-T_C\right)^{-\nu}$.\cite{Young_pc}
\begin{figure}
\includegraphics[width=3in]{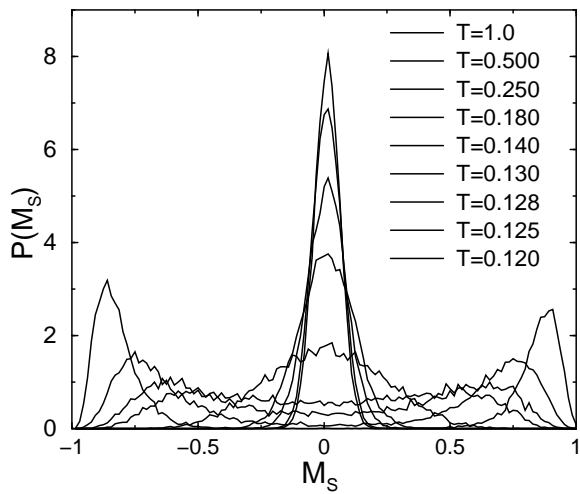}
\caption{The distribution $P(M_S)$ of the staggered occupation $M_S$ for
$L=8$ at $\sigma=0$ at various temperatures. The central peak is highest
at $T=1$ and gradually decreases as $T$ decreases. The two side peaks
begin to appear in the vicinity of $T_C$ and become more pronounced as
$T$ drops below $T_C$. There is no temperature where 3 peaks are present,
indicating that the transition is not first order.
The data was the result of averaging over 35 runs.
}
\label{fig:stagmag_sigma=0}
\end{figure}
Figure \ref{fig:stagmag_sigma=0} shows the distribution $P(M_s)$ of the
staggered occupation at various temperatures. Notice that in the
vicinity of the melting temperature there are only two symmetrical side
peaks. This implies that the transition is a second order phase transition.
Furthermore we find that the value $M_{s,max}$, 
where $P(M_s)$ has a maximum, decreases with increasing system size at $T_C$
and goes as $M_{s,max}\sim L^{-0.6}$. 
This is also consistent with a second order transition.
In the vicinity of the phase transition where $P(M_s)$ has 2 peaks,
we can define the width $M_{s,width}$ of the distribution
as the nonzero value of $M_{s}$ where $P(M_{s,width})=P(M_s=0)$.
We find at $T_C$ that $M_{s,width}$ is linear in $L$ and can be fit to the
form $M_{s,width}=A-mL$ where $A$ and $m$ are constants that are
temperature dependent. At $T=0.128$ which is close to $T_C$, $A=1.1$ and $m=0.027$.
Notice that $M_{s,width}$ does not appear to follow the form 
$M_{s,width}\sim L^{-\beta/\nu}$, 
but we would need more than 3 values of $L$ to accurately determine if there is a discrepancy
with the scaling form. 

First order phase transitions are often characterized by hysteresis upon
heating and cooling. We have looked for hysteresis by cooling and then heating
the system, and examining the resulting curves of $g$ versus $T$ 
as well as the specific
heat $C_V$ versus $T$. We find no hysteresis which is 
further evidence against a first order phase transition. 

To summarize, the ordered case ($\sigma=0$) undergoes a second order phase 
transition as a function of temperature.

\subsection{Critical Behavior}
\begin{figure}
\includegraphics[width=3in]{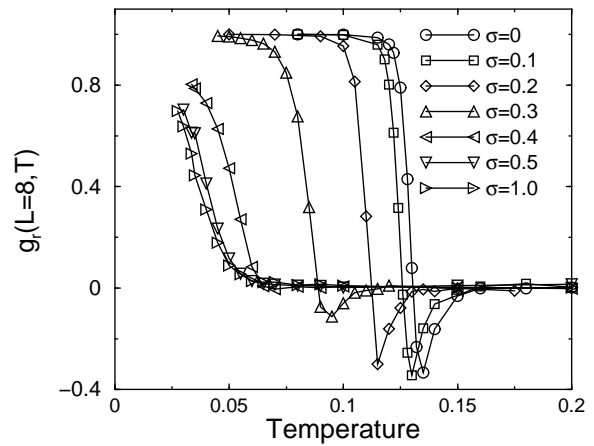}
\caption{$g_r(L=8,T)$ vs. $T$ for $\sigma=0$ (45 runs), $\sigma=0.1$ (10 runs),
$\sigma=0.2$ (5 runs), $\sigma=0.3$ (15 runs), $\sigma=0.4$ (115 runs),
$\sigma=0.5$ (45 runs), and $\sigma=1$ (108 runs). 
($g_t(L=8,T)$ vs. $T$ is virtually identical.) The number of runs in parentheses
is the number of runs that were averaged to obtain the data. The solid
lines are guides to the eye.}
\label{fig:gr_allsigmas}
\end{figure}

\begin{figure}
\includegraphics[width=3in]{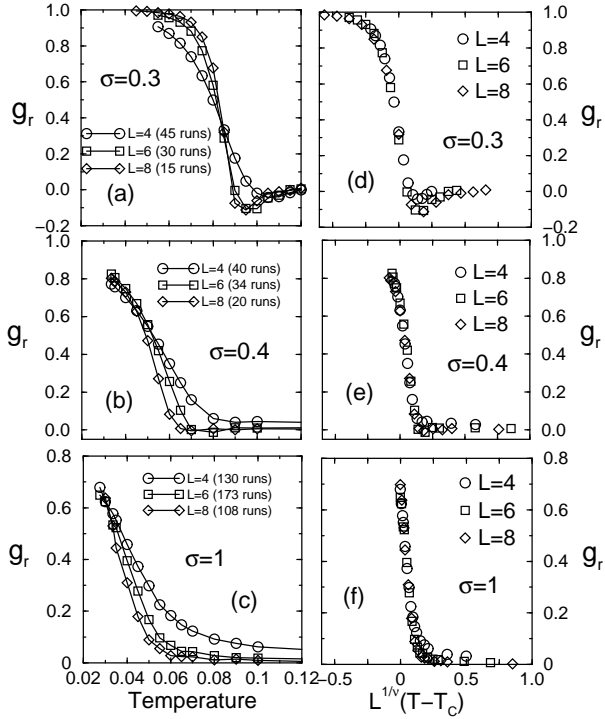}
\caption{(a)--(c) $g_r(L,T)$ versus $T$ for $\sigma=0.3$, 0.4, and 1.0
at $L=4$, 6, and 8. The solid lines are guides to the eye.
($g_r(L=8,T)$ vs. $T$ is virtually identical.) The number of runs 
in parentheses is the number of runs that were averaged to obtain the data. 
(d) $g(L,T)$ for $\sigma=0.3$ scaled using
$\hat{g}\left(L^{1/\nu}\left(T-T_C\right)\right)$ with $T_C=0.085\pm0.005$ and
$\nu=0.71\pm 0.1$. (e) $g(L,T)$ for $\sigma=0.4$ scaled using
$\hat{g}\left(L^{1/\nu}\left(T-T_C\right)\right)$ 
with $T_C=0.045\pm0.01$ and $\nu=1.05\pm 0.1$. 
(f) $g(L,T)$ for $\sigma=1$ scaled using
$\hat{g}\left(L^{1/\nu}\left(T-T_C\right)\right)$ with $T_C=0.028\pm 0.01$ and
$\nu=1.30\pm 0.2$.
}
\label{fig:gvsT_allL}
\end{figure}

\begin{figure}
\includegraphics[width=3in]{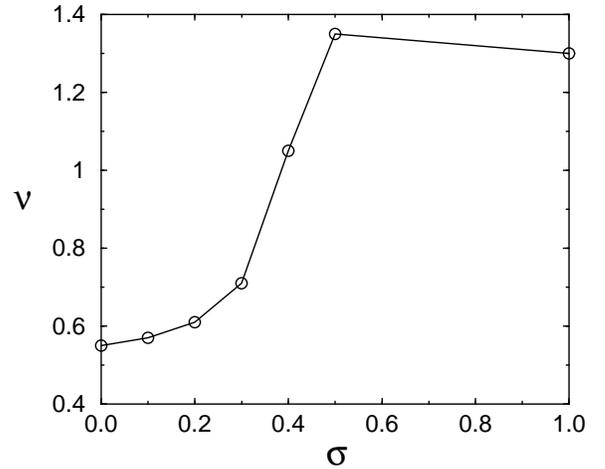}
\caption{The critical exponent $\nu$ versus the disorder $\sigma$.
The solid line is a guide to the eye.}
\label{fig:nu_vs_sigma}
\end{figure}

\begin{figure}
\includegraphics[width=3in]{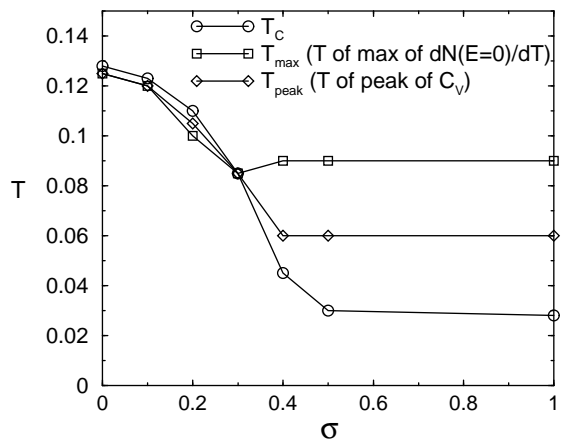}
\caption{Transition temperature $T_C$ vs. $\sigma$ ($\circ$), temperature $T_{max}$ 
of the maximum of $dN(E=0)/dT$ vs. $\sigma$ ($\Box$), and the temperature $T_{peak}$
of the maximum in the specific heat vs. $\sigma$ ($\Diamond$). 
The solid lines are guides to the eye.}
\label{fig:Ts_vs_sigma}
\end{figure}

We have determined the critical exponent $\nu$ and the transition 
temperature $T_C$ as a function of the disorder $\sigma$ through the 
finite size scaling of $g(L,T)$.\cite{Bhatt88,Grannan93} In Figure
\ref{fig:gr_allsigmas} we plot $g(L=8,T)$ versus $T$ for various values
of $\sigma$. 
Notice that the transition region moves to lower temperatures
with increasing disorder. This reflects the decrease in $T_C$ with increasing
$\sigma$. The transition temperature corresponds to the temperature
where the curves of $g(L,T)$ versus $T$ for all sizes cross. Examples
are shown in Figure \ref{fig:gr_sigma=0} and Figure \ref{fig:gvsT_allL}. 
To more accurately determine
$T_C$, we use the scaling hypothesis to collapse the data for a given  
value of $\sigma$ onto a single curve as shown in Figure \ref{fig:gvsT_allL}.
$T_C$ and $\nu$ are used as adjustable parameters to collapse the
data. The values of $\nu$ and $T_C$ at various values of $\sigma$
are given in table \ref{tab:table1}.
We can estimate the errors in the critical temperature and
the critical exponent $\nu$ by how well the curves can be made to 
collapse. The errors given in the
table also include our estimate of the effects of aging. In other words,
the error bars include our estimate of how the values might change
if we were to run longer at low temperatures or cool more slowly.
In Figures \ref{fig:nu_vs_sigma} 
and \ref{fig:Ts_vs_sigma} we plot $T_C$ and $\nu$ versus $\sigma$.

\begin{table}
\caption{\label{tab:table1} The values of $T_C$ and $\nu$ for different
valuse of $\sigma$.}
\begin{ruledtabular}
\begin{tabular}{ccc}
$\sigma$ & $T_C$ & $\nu$\\
\hline
0.0 & $0.128\pm 0.005$ & $0.55\pm 0.1$\\
0.1 & $0.123\pm 0.005$ & $0.57\pm 0.1$\\
0.2 & $0.110\pm 0.005$ & $0.61\pm 0.1$\\
0.3 & $0.085\pm 0.005$ & $0.71\pm 0.1$\\
0.4 & $0.045\pm 0.01$ & $1.05\pm 0.2$\\
0.5 & $0.030\pm 0.01$ & $1.35\pm 0.2$\\
1.0 & $0.028\pm 0.01$ & $1.30\pm 0.2$\\
\end{tabular}
\end{ruledtabular}
\end{table}

We can see that $\nu$ increases from $\nu=0.55\pm 0.1$ at $\sigma=0$ to
$\nu=1.30\pm 0.2$ at $\sigma=1$. The value of $\nu$ in the
ordered case ($\sigma=0$) lies between the classical value ($\nu=0.5$)
and the value for the ordered short ranged Ising model 
($\nu=0.63$) \cite{Fisher94}. Within the error bars, our value
is consistent with both universality classes, and therefore cannot
differentiate between them.
M{\"o}bius and R{\"o}{\ss}ler \cite{Mobius03} studied a half--filled
system on a cubic lattice with Coulomb interactions and found
$\nu=0.635(10)$ which agrees with the value for the Ising model.
Our simulations differ from those of M{\"o}bius and 
R{\"o}{\ss}ler in that we used the Ewald 
summation to take into account the fact that 
the Coulomb interaction extends beyond the size of the system 
while they did not. As we mentioned earlier, the order-disorder
transition in the LRPM model on a simple cubic lattice belongs 
to the Ising universality class \cite{privCommCiach}. 
The completely filled LRPM model with ionic density $\rho=1$ is
equivalent to our $\sigma=0$ case. In addition Luijten {\it et al.} did
Monte Carlo studies of the restricted
primitive model (RPM) which has equal numbers of oppositely
charged ions with equal diameters and with Coulomb interactions in
three dimensions \cite{Luijten02b}. 
These grand canonical simulations of the RPM used a finely
discretized lattice where the ionic diameters were 5 times
larger than the lattice spacing, and they found the Ising
of the critical exponent $\nu=0.63(3)$.

In the disordered case ($\sigma=1$) our value for $\nu=1.3\pm 0.2$ 
differs from the value of $\nu=0.75^{+0.2}_{-0.1}$ 
obtained earlier \cite{Grannan93}. Again this is probably due
to the fact that we used Ewald summation whereas the previous work
did not. In addition we were able to do longer runs at low
temperatures than the previous work. 

The transition
temperature decreases from $T_C=0.128\pm 0.005$ at $\sigma=0$ to
$T_C=0.028\pm 0.01$ at $\sigma=1$. The value of $T_C=0.128$ at $\sigma=0$
is consistent with the temperature of the peak in the specific heat
found previously by M{\"o}bius and R{\"o}{\ss}ler \cite{Mobius03}.
Within the error
the value of $T_C=0.028\pm 0.01$ at $\sigma=1$ is consistent with
the previous value of $T_C=0.043^{+0.003}_{-0.006}$ found by
Grannan and Yu \cite{Grannan93}. 

It is interesting that $T_C$ is much lower than the characteristic energies
of the system which are of order unity. This is especially true for large values
of the disorder. The reason for this was given by Grannan and Yu \cite{Grannan93}
and is as follows. At the temperatures of our simulations, nearby pairs of 
sites will with high probability consist of an occupied and an unoccupied site.
Since these strongly coupled pairs of sites are close together, they are 
guaranteed to have small dipole moments. Therefore, they will interact
weakly with the rest of the system, remaining active down to temperatures
much lower than the bare interaction energy.

\subsection{Specific Heat}
\begin{figure}
\includegraphics[width=3in]{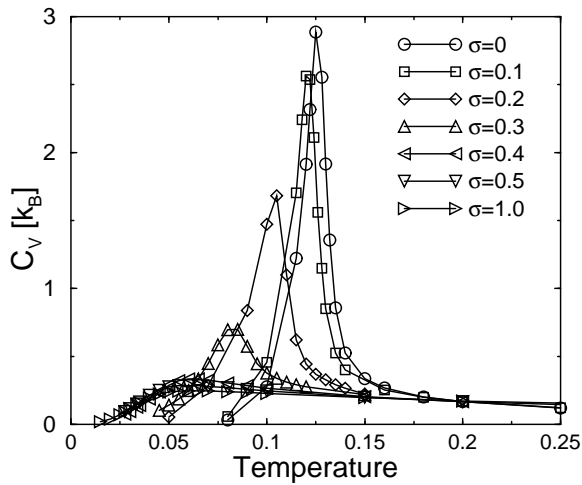}
\caption{The specific heat $C_V$ versus $T$ in units of $k_B$ for $L=8$ for
$\sigma=0$ (45 runs), $\sigma=0.1$ (10 runs), $\sigma=0.2$ (5 runs),
$\sigma=0.3$ (15 runs), $\sigma=0.4$ (95 runs), $\sigma=0.5$ (45 runs), and
$\sigma=1$ (108 runs). The number of runs averaged over is indicated in
parentheses. The solid lines are guides to the eye.
}
\label{fig:spht_vs_T_allsigma_L=8}
\end{figure}
In Figure \ref{fig:spht_vs_T_allsigma_L=8} we plot the specific heat versus
temperature for various values of $\sigma$ for $L=8$. We see that in the
ordered case ($\sigma=0$) $C_V$ exhibits a sharp peak centered at $T_C$.
As the disorder increases, the peak broadens and eventually becomes
a broad bump with a maximum at a temperature above $T_C$. For example,
for $\sigma=1$, $C_V$ has a maximum at $T=0.07$ whereas $T_C=0.028$.
In Figure \ref{fig:Ts_vs_sigma} we compare the temperature $T_{peak}$ of
the maximum in the specific heat with $T_C$ for various values of $\sigma$.
We see that $T_{peak}$ matches well with $T_C$ for $\sigma \leq 0.3$. For
larger values of $\sigma$, $T_{peak}>T_C$.
Spin glasses also have a maximum in their specific heat at a temperature
above the spin glass transition temperature \cite{Binder86}. 
For the three dimensional Coulomb glass where the disorder is in 
the onsite energy rather than
in the positions of the sites, M{\"o}bius {\it et al.} found that
as the width in the distribution of onsite energies increased,
the temperature $T_{peak}$ of the maximum in the $C_V$ also increased
\cite{Mobius97,Mobius01}.  However, in the cases of onsite disorder that they 
considered, the maximum does not signify a transition since the existence
of a phase transition in the presence of onsite disorder has 
not been established. 
The maximum in $C_V$ must be present since $C_V(T)$ goes to zero
at the extremes $T\rightarrow 0$ and $T\rightarrow\infty$, implying
that there must be a maximum in between these extremes \cite{Mobius04}. 
Furthermore, even without Coulomb interactions but with a large amount
of onsite disorder, there would be a maximum in the specific
heat consisting of a superposition of the Schottky specific heats of
two level systems with randomly distributed excitation energies
\cite{Mobius04}.

\begin{figure}
\includegraphics[width=3in]{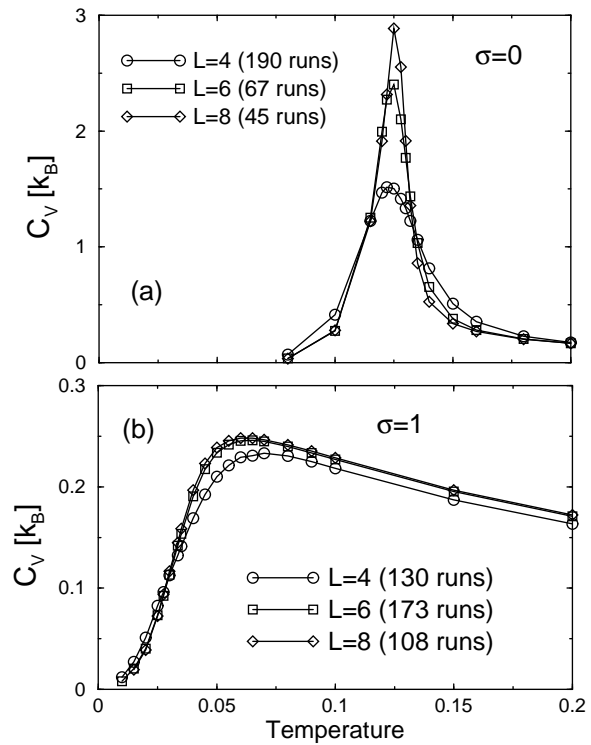}
\caption{The specific heat $C_V$ versus $T$ in units of $k_B$ for $L=4$,
6, and 8 for (a) $\sigma=0$ and (b) $\sigma=1$.  The specific heat
is calculated from the variance in the energy fluctuations. The specific
heat calculated from the derivative of the energy with respect to temperature
is similar.  The number of runs averaged over is indicated in
parentheses. The solid lines are guides to the eye.
}
\label{fig:spht_vs_T_allL}
\end{figure}
To show the size dependence of the specific heat, in Figure
\ref{fig:spht_vs_T_allL} we plot $C_V$ versus $T$ for different system 
sizes at $\sigma=0$ and at $\sigma=1$. In the ordered case the specific 
heat peak becomes sharper as $L$ increases while in the disordered
case, the broad bump is only weakly dependent on system size.

\subsection{Single Particle Density of States $N(E)$}
In a Coulomb glass the long range Coulomb interactions between 
localized electrons produce a Coulomb gap in the single particle 
density of states that is centered at the Fermi energy 
\cite{Pollak70,Efros75,efrosbook}. The Coulomb gap makes the ground
state stable with respect to single electron hops. 
The ordered case also has a gap but for a somewhat different reason.
In the ground state of the ordered case where there is an FCC lattice, the
potential energy or local field is the same for each occupied site. The local field is
equal and opposite for the unoccupied sites. This leads to an $N(E)$ with
two delta functions symmetrically placed about $E=0$. 
In finite size systems at finite temperatures these delta functions
broaden into finite height peaks due to thermal fluctuations and the formation
of ordered domains.

\begin{figure}
\includegraphics[width=3in]{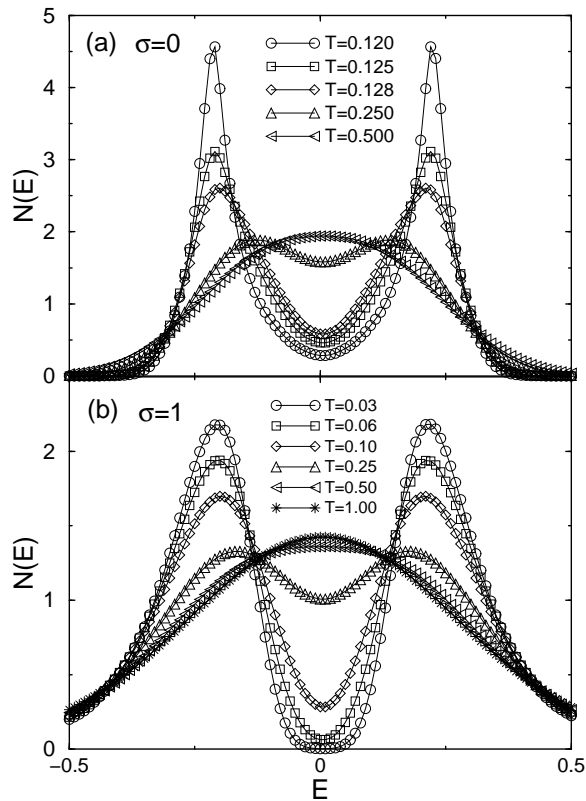}
\caption{$N(E)$ versus $E$ for $L=8$ at various temperatures.
The solid lines are guides to the eye.
(a) $\sigma=0$. The data was averaged over 45 runs.
(b) $\sigma=1$. The 3 lowest temperatures were averaged over 108 runs
and the 3 highest temperatures were averaged over 16 runs. 
}
\label{fig:nE_sigma=01}
\end{figure}
\begin{figure}
\includegraphics[width=3in]{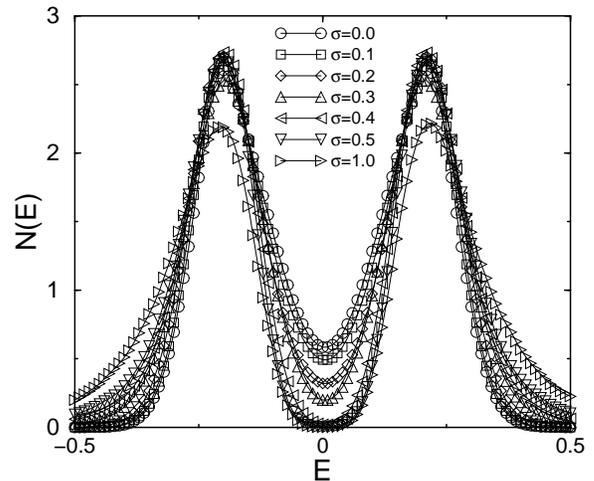}
\caption{$N(E)$ versus $E$ for $L=8$ for various values of $\sigma$
at temperatures in the vicinity of $T_C$. Shown are $\sigma=0$ ($T=0.128$, 45 runs),
$\sigma=0.1$ ($T=0.122$, 10 runs), $\sigma=0.2$ ($T=0.105$, 5 runs), 
$\sigma=0.3$ ($T=0.085$, 15 runs),
$\sigma=0.4$ ($T=0.045$, 115 runs), $\sigma=0.5$ ($T=0.030$, 45 runs), 
and $\sigma=1$ ($T=0.0275$, 108 runs).
The temperatures and the number of runs averaged over is indicated in
parentheses. The solid lines are guides to the eye.
}
\label{fig:nE_allsigma}
\end{figure}

In Figure \ref{fig:nE_sigma=01} we show the density of states $N(E)$
for single particle excitations at various temperatures for $\sigma=0$
and for $\sigma=1$. Because of strong electron--electron correlations,
the density of states at zero energy starts to decrease at about $2T_C$ 
in the ordered case ($\sigma=0$) but at a temperature about an order 
of magnitude above $T_C$ in the strongly disordered case ($\sigma=1$).
In Figure \ref{fig:nE_allsigma} we show $N(E)$ at or near $T_C$ for various
values of $\sigma$ for L=8. We see that at $T_C$ the gap appears 
nearly fully formed for
$\sigma=1$ but not for $\sigma=0$. In the ordered case the finite density of
states at $E=0$ is possibly due to domains. 

As we can see from the figures, at finite temperatures the gap 
in the density of states is partially filled, and the density of 
states does not vanish at the Fermi energy. This has been seen in
previous simulations
\cite{Levin87,Grannan93,Li94,Mogilyanskii89,Vojta93,Sarvestani95}.
Tunneling measurements of the Coulomb gap have also seen that it
fills in with increasing temperature \cite{Lee99,Sandow01}.
The exact form of $N(E,T)$ is not known, but for strong disorder
some have argued \cite{Levin87,Vojta93,Li94} that 
its low temperature asymptotic behavior is described by $N(E=0,T)\sim T^{d-1}$.
However, some simulations \cite{Sarvestani95} have found
a stronger temperature dependence, i.e., $N(E=0,T)\sim T^{\lambda}$ with
$\lambda > (d-1)$. For $d=2$, Sarvestani {\it et al.} \cite{Sarvestani95}
found $\lambda=1.75\pm 0.1$, and for $d=3$, $\lambda=2.7\pm 0.1$.

\begin{figure}
\includegraphics[width=3in]{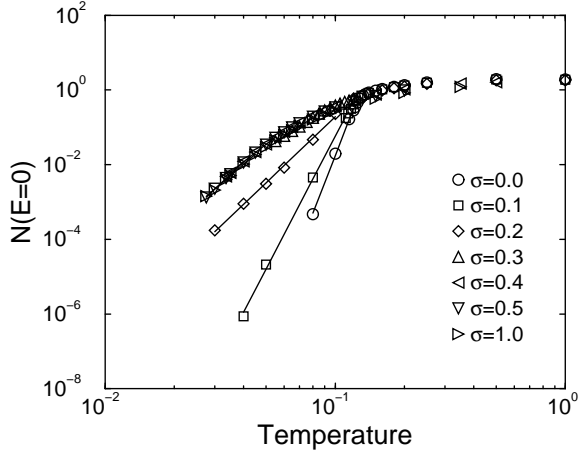}
\caption{Log--log plot of $N(E=0)$ versus $T$ for $L=8$ for $\sigma=0$ (45 runs),
$\sigma=0.1$ (15 runs), $\sigma=0.2$ (10 runs), $\sigma=0.3$ (10 runs),
$\sigma=0.4$ (115 runs), $\sigma=0.5$ (45 runs), and $\sigma=1$ (108 runs).
The number of runs averaged over is indicated in
parentheses. The solid lines are power law fits to the data at
low temperatures.
}
\label{fig:nE=0vsT_log}
\end{figure}

In Figure \ref{fig:nE=0vsT_log} we show our results in
a log--log plot of $N(E=0,T)$ for various 
values of $\sigma$ at $L=8$. 

\begin{figure}
\includegraphics[width=3in]{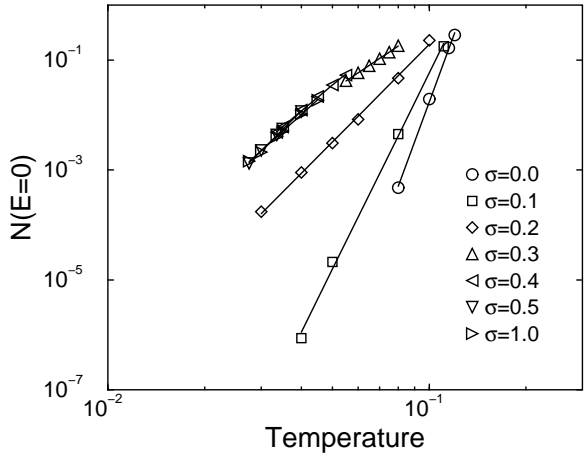}
\caption{Log--log plot of $N(E=0)$ versus $T$ at low temperature
for $L=8$ for $\sigma=0$ (45 runs),
$\sigma=0.1$ (15 runs), $\sigma=0.2$ (10 runs), $\sigma=0.3$ (10 runs),
$\sigma=0.4$ (115 runs), $\sigma=0.5$ (45 runs), and $\sigma=1$ (108 runs).
The number of runs averaged over is indicated in
parentheses. The solid lines are power law fits to the form
$N(E=0,T)\sim T^{\lambda}$.
}
\label{fig:nE=0vslowT_log_fit}
\end{figure}

\begin{figure}
\includegraphics[width=3in]{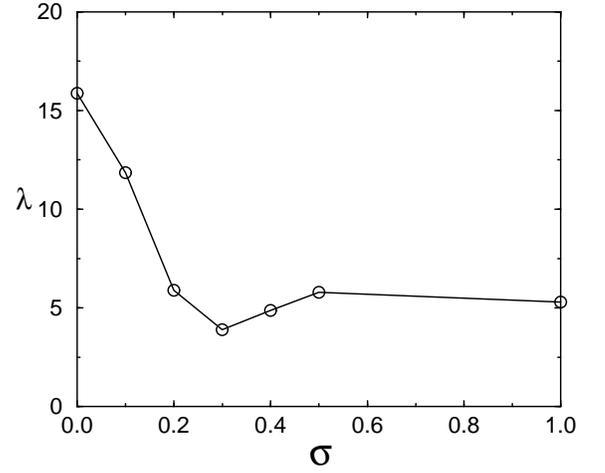}
\caption{The power $\lambda$ versus $\sigma$ for $L=8$.
The solid line is a guide to the eye.
}
\label{fig:powerNE_vs_sigma}
\end{figure}

At low temperatures the curves are quite straight on a log--log plot. So
we can fit the low temperature part of these curves to a power law form
$N(E=0,T)\sim T^{\lambda}$. The fits are shown as solid lines in
Figure \ref{fig:nE=0vsT_log} and in Figure \ref{fig:nE=0vslowT_log_fit}.
We plot $\lambda$ as a function of $\sigma$
in Figure \ref{fig:powerNE_vs_sigma}. We find that $\lambda$ varies between 
3 to 16 and is always greater than $d-1=2$ since $d=3$. Even in the 
case of uniform disorder (uniform random), $\lambda=4.8$. 
The large value of $\lambda$ for $\sigma=0$ is not entirely surprising
since Figure \ref{fig:nE_allsigma} shows that $N(E=0,T)$ is larger 
for $\sigma=0$ than for any other value of the disorder in the vicinity
of the transition temperature. Since $N(E=0,T)$ goes to zero as the temperature
goes to zero, $N(E=0,T)$ has the farthest to go for $\sigma=0$. Even
though $T_C$ is largest for $\sigma=0$, the ratio $N(E=0,T)/T_C$ is largest
for the case of no disorder, and so
it is consistent that the exponent $\lambda$ is largest for the case of
no disorder.

\begin{figure}
\includegraphics[width=3in]{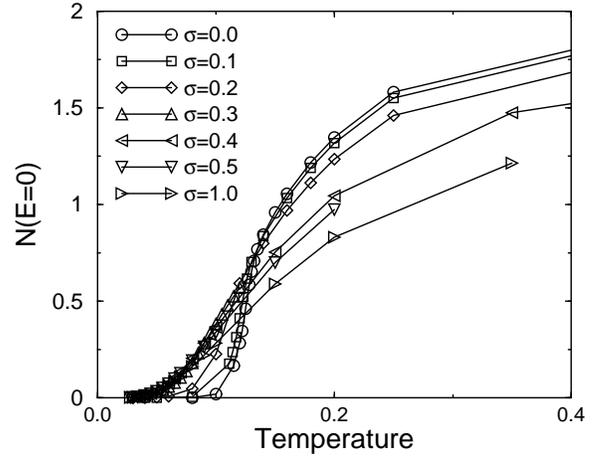}
\caption{Linear plot of $N(E=0)$ versus $T$ for $L=8$. Data is the
same as is shown in Fig. \ref{fig:nE=0vsT_log}. The solid lines are
guides to the eye.
}
\label{fig:nE=0vsT}
\end{figure}

We plot the data from Fig. \ref{fig:nE=0vsT_log} on a linear plot in 
Figure \ref{fig:nE=0vsT} where we see S--shaped curves.
We can see that $N(E=0,T)$ rises much more steeply for small values of
disorder than for large values of disorder. The steepest part rise for 
the ordered cases ($\sigma\leq 0.3$) occurs approximately at $T_C$. 
We can quantify this by taking a derivative $dN(E=0)/dT$ that can be
approximated by a finite difference:
\begin{equation}
\left. \frac{dN(E=0)}{dT}\right|_{T=T_{i}}\approx
\frac{N_{i+1}(E=0)-N_{i}(E=0)}{T_{i+1}-T_{i}} 
\end{equation}
The result is shown in Figure \ref{fig:Ts_vs_sigma}
where we compare $T_C$ with the temperature $T_{max}$ 
where $dN(E=0)/dT$ is a maximum.
We see that $T_{max}$ follows $T_C$ for $\sigma\leq 0.3$ but lies above $T_C$
for larger values of the disorder $\sigma$.

Efros and Shklovskii \cite{Efros75,efrosbook} argued that at 
$T=0$ the Coulomb gap in the single particle density of states 
of a fully disordered system should scale as 
$N(E)\sim|E-E_F|^{\delta}$ where $\delta=d-1$ 
and $d$ is the dimension of the system. Some subsequent work
\cite{Levin87,Vojta93,Baranovskii79} has supported this form
for the density of states,
though some simulations \cite{Davies84,Li94,Sarvestani95,Mobius92}
have found a steeper energy dependence, i.e., $d > \delta > (d-1)$
in two \cite{Davies84,Li94,Sarvestani95,Mobius92} and three
\cite{Li94,Sarvestani95,Mobius92} dimensions.
Efros \cite{Efros76} included two--electron transitions in calculating
the density of states of a Coulomb glass and proposed
the exponential form
$N(E)\sim\exp\left[-|E_o/(E-E_F)|^{1/2}\right]$ where $E_o$ is a constant.
The physical reason for such a sharp gap is the formation of polarons
in which an occupied site tends to have unoccupied sites nearby
and vice--versa. Some simulations \cite{Davies84}
have found support for this exponential form, while others
\cite{Baranovskii79,Mobius92} have not.

\begin{figure}
\includegraphics[width=3in]{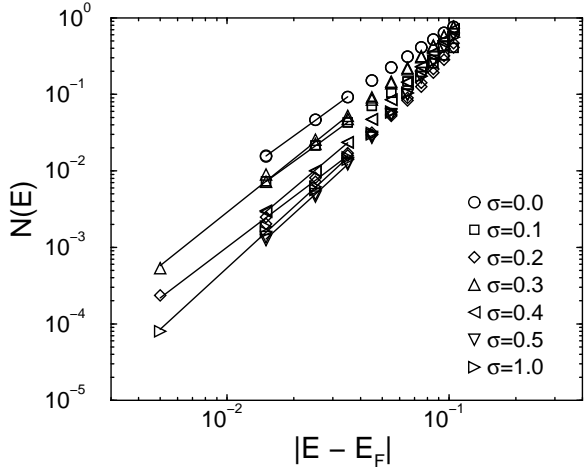}
\caption{Log-log plot of $N(E)$ versus $|E-E_F|$ for $L=8$ for 
$\sigma=0$ ($T=0.120$, 45 runs), $\sigma=0.1$ ($T=0.111$, 15 runs),
$\sigma=0.2$ ($T=0.08$, 20 runs), $\sigma=0.3$ ($T=0.055$, 15 runs),
$\sigma=0.4$ ($T=0.0335$, 115 runs), $\sigma=0.5$ ($T=0.0275$, 45 runs),
$\sigma=1.0$ ($T=0.0300$, 108 runs).  The temperatues are all below
$T_C$. The solid lines are
fits to a power law $[N(E)-N(0)]\sim|E-E_F|^{\delta}$ for values of $E$ very
close to the Fermi energy $E_F$, i.e., $|E-E_F|<0.04$. 
The plots include $N(E)$ values for $E$ above and below $E_F$.
}
\label{fig:powerlaw_NE_vs_E}
\end{figure}

\begin{figure}
\includegraphics[width=3in]{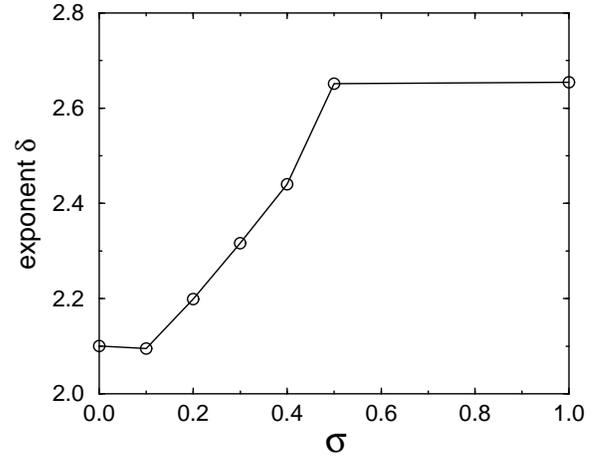}
\caption{The exponent $\delta$ versus $\sigma$ from the fits
to the power law $[N(E)-N(0)]\sim|E-E_F|^{\delta}$
in Fig.~\ref{fig:powerlaw_NE_vs_E}. The solid line is a guide to the eye.
The error in $\delta$ is approximately $\pm 0.2$ due to finite
size and finite temperature effects described in the text.
}
\label{fig:power_delta_vs_sigma}
\end{figure}

According to the theory \cite{Efros75,efrosbook}, $N(E)\sim|E-E_F|^{d-1}$
in the limit $E\rightarrow E_F$. In Figure \ref{fig:powerlaw_NE_vs_E} we
plot our data for $[N(E)-N(E_F)]$ versus $|E-E_F|$ on a log--log plot. (Since
we are at finite temperatures, we have subtracted off $N(E_F)$.) We fit
the low energy data in the vicinity of the Fermi energy $E_F$ to 
a power law of the form $N(E)\sim|E-E_F|^{\delta}$ for various values
of $\sigma$ at temperatures below $T_C$. For the case of strong
disorder ($\sigma=1$), we find $\delta=2.65\pm 0.2$ which agrees with
the previous values of $\delta=2.6\pm 0.2$ found by M{\"o}bius {\it et al.}
\cite{Mobius92} and $\delta=2.7\pm 0.1$ found by Sarvestani {\it et al.} 
\cite{Sarvestani95}. It disagrees with the value of $\delta=d-1=2$ predicted
by Efros and Shklovskii \cite{Efros75,efrosbook} and with the value
$\delta=2.38$ found by Li and Phillips \cite{Li94}. In the case
of no disorder, the curvature is very close to quadratic
and we find $\delta=2.1\pm 0.2$. In Figure \ref{fig:power_delta_vs_sigma}
we plot the exponent $\delta$ versus the disorder $\sigma$. We
see that $\delta$ increases and then saturates with increasing
disorder. The estimated error of $\pm 0.2$ in $\delta$ does not
come from the fit to the data, so much as from the fact that
the finite temperature affects low energies $E\stackrel{<}{\sim} kT$.
There are also finite size effects \cite{Mobius92} that affect low
energies $E\stackrel{<}{\sim} 1/2L$, though finite size effects for
$L\ge 6$ are quite small (less than 1\%). 

\begin{figure}
\includegraphics[width=3in]{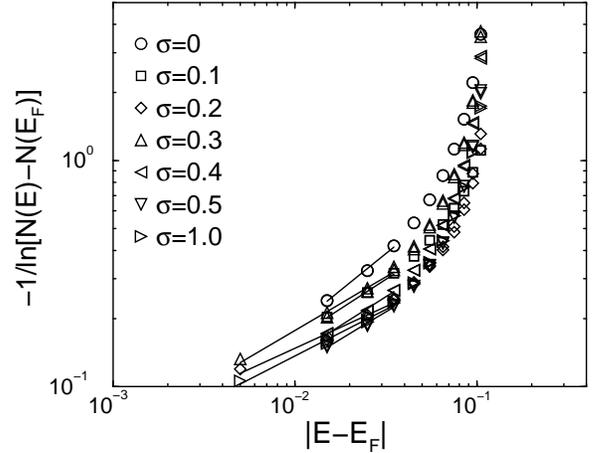}
\caption{Log--log plot of $-1/\ln[N(E)-N(E_F)]$ versus $|E-E_F|$ for 
$L=8$ for $\sigma=0$ ($T=0.120$, 45 runs), $\sigma=0.1$ ($T=0.111$, 15 runs),
$\sigma=0.2$ ($T=0.08$, 20 runs), $\sigma=0.3$ ($T=0.055$, 15 runs),
$\sigma=0.4$ ($T=0.0335$, 115 runs), $\sigma=0.5$ ($T=0.0275$, 45 runs),
$\sigma=1.0$ ($T=0.0300$, 108 runs).  The temperatues are all below
$T_C$. The solid lines are
fits to $-1/\ln[N(E)-N(E_F)]\sim|E-E_F|^{\gamma}$ for values of $E$ very
close to the Fermi energy $E_F$, i.e., $|E-E_F|<0.04$. The slope of each
line gives $\gamma$. 
The plots include $N(E)$ values for $E$ above and below $E_F$.
}
\label{fig:stretchedExpFit}
\end{figure}

\begin{figure}
\includegraphics[width=3in]{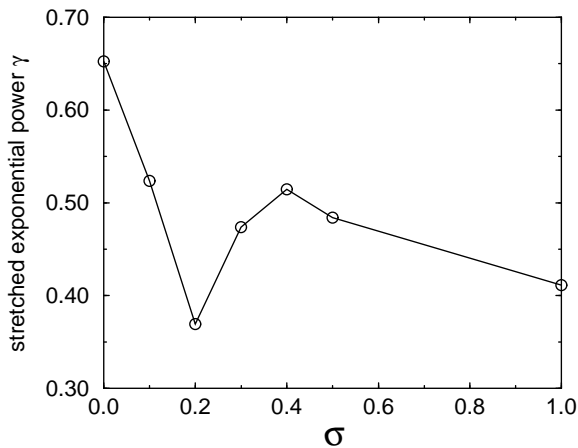}
\caption{The exponent $\gamma$ versus $\sigma$ from the fits
to the exponential form 
$[N(E)-N(E_F)]\sim\exp\left|-E_o/(E-E_F)\right|^{\gamma}$
in Fig.~\ref{fig:stretchedExpFit}. The solid line is a guide to the eye.
}
\label{fig:stretchedExpPower_vs_sigma}
\end{figure}

We have checked to see if our data provides evidence for the
exponential form 
$N(E)\sim\exp\left[-|E_o/(E-E_F)|^{1/2}\right]$ proposed
by Efros \cite{Efros76}. In Figure \ref{fig:stretchedExpFit}
we show a log-log plot of $-1/\ln[N(E)-N(E_F)]$ versus $|E-E_F|$ for
various values of $\sigma$.
(Since we are at finite temperatures, we have subtracted off $N(E_F)$.)
If $N(E)\sim\exp\left[-|E_o/(E-E_F)|^{1/2}\right]$ were a good description
of the density of states, then the curves in 
Figure \ref{fig:stretchedExpFit} would be straight lines
with slopes of 1/2. Since the exponential form presumably only describes
the density of states in the vicinity of $E_F$, we have fit lines
through the points corresponding to $|E|<0.04$ assuming the more
general form $[N(E)-N(E_F)]\sim\exp-\left|E_o/(E-E_F)\right|^{\gamma}$. 
The slope of the lines in Figure \ref{fig:stretchedExpFit} correspond
to the exponent $\gamma$. We plot $\gamma$ versus $\sigma$ in 
Figure \ref{fig:stretchedExpPower_vs_sigma}. The values of $\gamma$
fluctuate around $1/2$, but the large curvature of the trajectories
in Figure \ref{fig:stretchedExpFit} do not lend strong support
to the exponential form of the density of states.

\begin{figure}
\includegraphics[width=3in]{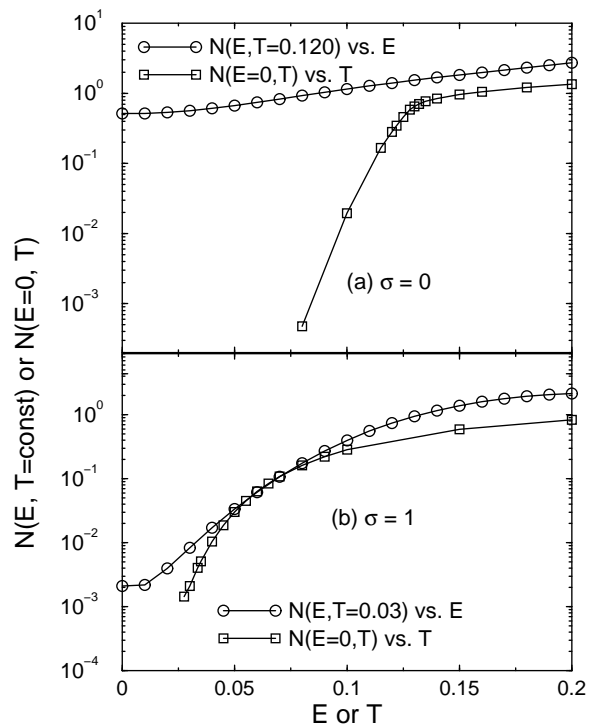}
\caption{(a) $N(E=0,T)$ vs.~$T$, and $N(E,T=0.120)$ vs.~$E$ for
$\sigma=0$. The data is averaged over 45 runs.
(b) $N(E=0,T)$ vs.~$T$, and $N(E,T=0.0275)$ vs.~$E$ for $\sigma=1$.
The data is averaged over 108 runs. The solid lines are guides to the eye.
}
\label{fig:NE_vs_E_T}
\end{figure}

Analytical theories \cite{Vojta93,Mogilyanskii89} of the Coulomb glass
predict that the finite temperature density of states
$N(E=0,T)$ at the Fermi energy ($E_F=0$) should be 
proportional to the zero temperature
density of states $N(E,T=0)$ at an energy $E=k_BT$, i.e.,
$N(E=0,T)\sim N(E,T=0)$ with $|E-E_F|=k_BT$. This has been supported by
Coulomb glass simulations \cite{Sarvestani95}. We tested this relation by plotting
$N(E=0,T)$ versus $T$, and $N(E,T=T_o)$ versus $E$ on the same graph,
where $T_o$ is the lowest temperature at which we were able to equilibrate
the system. We show our results in Figure \ref{fig:NE_vs_E_T}
for $\sigma=0$ and 1. The hypothesis seems to work for a limited
range of energies between $k_BT_o$ and the width of the Coulomb gap. It also
appears to be more applicable for high disorder ($\sigma=1$) than for the
case of no disorder ($\sigma=0$).

\subsection{Staggered Occupation}
\begin{figure}
\includegraphics[width=3in]{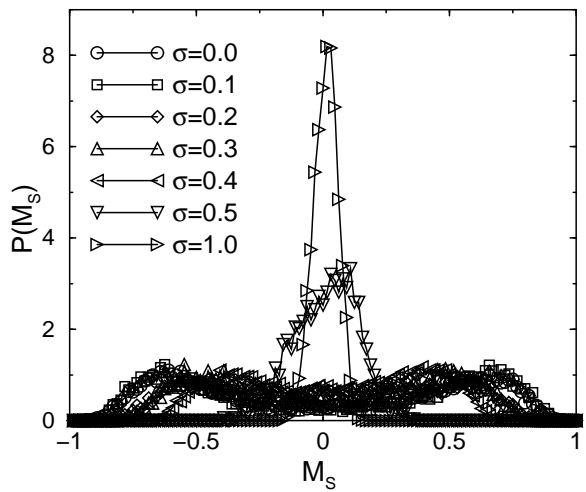}
\caption{Staggered occupation distribution for $L=8$ for various
values of $\sigma$ in the vicinity of $T_C$. $\sigma=0$ (35 runs,
$T=0.128$), $\sigma=0.1$ (15 runs, $T=0.123$), $\sigma=0.2$ 
(10 runs, $T=0.110$), $\sigma=0.3$ (10 runs, $T=0.085$),
$\sigma=0.4$ (40 runs, $T=0.045$), $\sigma=0.5$ (10 runs, $T=0.03$), 
and $\sigma=1$ (10 runs, $T=0.03$). The number of runs averaged
over is indicated in parentheses. The solid lines are guides to the eye.
}
\label{fig:stag_mag_allsigmas_L=8_Tc}
\end{figure}

\begin{figure}
\includegraphics[width=3in]{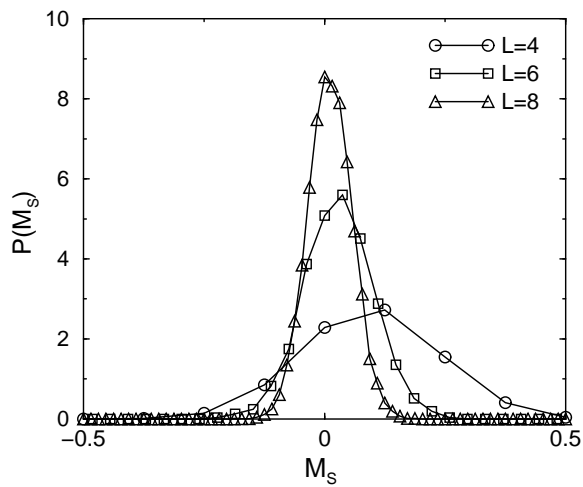}
\caption{Staggered occupation distribution for $\sigma=1$
at $T=1$ for $L=4$, 6, and 8. The peak height increases
with increasing $L$. The data shown is the result of
averaging over 10 runs. The solid lines are guides to the eye.
}
\label{fig:stagmag_sigma=1_T=1_allL}
\end{figure}

We have studied the staggered occupation at various values of the
disorder. At high temperatures the distribution has a peak centered at
$M_s=0$ for all values of the disorder. At low temperatures the distribution 
broadens and has two peaks symmetrically placed about zero for the ordered case
and for small and moderate values of the disorder. For the strongly
disordered case $\sigma\geq 0.5$, the distribution has a peak centered 
at $M_s=0$ for
all values of the temperature where the system was able to attain
equilibrium in our simulations. 
This is what one would expect for a random system. 
These features are illustrated in Figure \ref{fig:stag_mag_allsigmas_L=8_Tc} 
which shows the staggered occupation for various values of $\sigma$ in
the vicinity of $T_C$.
As a function of system size, the high temperature peak in $P(M_s)$ becomes
sharper as $L$ increases for all values of $\sigma$.
An example is shown in Figure \ref{fig:stagmag_sigma=1_T=1_allL}.

\section{Summary}
We have performed a Monte Carlo study of a classical three dimensional
Coulomb system of electrons
in which we systematically increase the positional disorder
by introducing deviations from positions in a cubic lattice.
We start from a completely ordered system and gradually
transition to a Coulomb glass.
The phase transition as a function of temperature is second order for
all values of disorder.
We use finite size scaling to determine the transition temperature $T_C$ and
the critical exponent $\nu$. We find that $T_C$ decreases and that $\nu$
increases with increasing disorder. Both quantities saturate in
the limit of large disorder. The specific
heat peak value decreases and the peak broadens to a broad bump with
increasing disorder. A gap develops in
the single particle density of states for all values of $\sigma$. At
low temperatures $N(E=0)\sim T^{\lambda}$ where $\lambda > 3.8$
for all values of $\sigma$. 
At low temperatures and low energies near $E_F$, the density of
states can be fit to a power law form $N(E)\sim|E-E_F|^{\delta}$
where $d-1<\delta<d$ for all values of $\sigma$. $\delta$ increases
with increasing $\sigma$, starting at $\delta=2.1$ for $\sigma=0$
and saturating at $\delta=2.65$ for $\sigma=1$.
The distribution of the staggered occupation has
a single central peak at high temperature for all values of the
disorder. In the ordered cases ($\sigma\leq 0.4$) $P(M_s)$ develops
two peaks symmetrically placed on either side of $M_s=0$ in the
vicinity of the phase transition.

We thank Peter Young, Robijn Bruinsma, Alina Ciach, and Sylvain Grollau
for helpful discussions. We thank Joseph Snider for technical assistance.
This work was supported by DOE grants DE-FG03-00ER45843 and DE-FG02-04ER46107. 
Work done while CCY was visiting the Kavli
Institute for Theoretical Physics at the University of California,
Santa Barbara was supported in part by the National Science Foundation under
Grant No. PHY99-07949.


\begin{thebibliography}{10}

\bibitem{Mobius97}
A. M{\"o}bius and P. Thomas, Phys. Rev. B {\bf 55},  7460  (1997).

\bibitem{Mobius01}
A. M{\"o}bius, P. Thomas, J. Talamantes, and C.~J. Adkins, Phil. Mag. B {\bf
  81},  1105  (2001).

\bibitem{Grannan93}
E.~R. Grannan and C.~C. Yu, Phys. Rev. Lett. {\bf 71},  3335  (1993).

\bibitem{Davies82}
J.~H. Davies, P.~A. Lee, and T.~M. Rice, Phys. Rev. Lett. {\bf 49},  758
  (1982).

\bibitem{Davies84}
J.~H. Davies, P.~A. Lee, and T.~M. Rice, Phys. Rev. B {\bf 29},  4260  (1984).

\bibitem{Grunewald82}
M. Gr\"{u}newald, B. Pohlmann, L. Schweitzer, and D. Wurtz, J. Phys. C {\bf
  15},  L1153  (1982).

\bibitem{Vojta93b}
T. Vojta, J. Phys. A: Math. Gen. {\bf 26},  2883  (1993).

\bibitem{Pollak70}
M. Pollak, Discuss. Faraday Soc. {\bf 50},  13  (1970).

\bibitem{Efros75}
A.~L. Efros and B.~I. Shklovski\u{i}, J. Phys. C {\bf 8},  L49  (1975).

\bibitem{efrosbook}
B.~I. Shklovski\u{i} and A.~L. Efros, {\em Electronic Properties of Doped
  Semiconductors} (Spinger-Verlag, Berlin, 1984).

\bibitem{Levin87}
E.~I. Levin, V.~L. Nguyen, B.~I. Shklovski\u{i}, and A.~L. \'{E}fros, Sov.
  Phys. JETP {\bf 65},  842  (1987).

\bibitem{Li94}
Q. Li and P. Phillips, Phys. Rev. B {\bf 49},  10269  (1994).

\bibitem{Mogilyanskii89}
A.~A. Mogilyanski\u{i} and M.~E. Ra\u{i}kh, Sov. Phys. JETP {\bf 68},  1081
  (1989).

\bibitem{Vojta93}
T. Vojta, W. John, and M. Schreiber, J. Phys.: Condens. Matter {\bf 5},  4989
  (1993).

\bibitem{Sarvestani95}
M. Sarvestani, M. Schreiber, and T. Vojta, Phys. Rev. B {\bf 52},  R3820
  (1995).

\bibitem{Massey95}
J.~G. Massey and M. Lee, Phys. Rev. Lett. {\bf 75},  4266  (1995).

\bibitem{Massey96}
J.~G. Massey and M. Lee, Phys. Rev. Lett. {\bf 77},  3399  (1996).

\bibitem{Lee99}
M. Lee, J.~G. Massey, V.~L. Nguyen, and B.~I. Shklovskii, Phys. Rev. B {\bf
  60},  1582  (1999).

\bibitem{Sandow01}
B. Sandow {\it et~al.}, Phys. Rev. Lett. {\bf 86},  1845  (2001).

\bibitem{Ciach02}
A. Ciach and G. Stell, Physica A {\bf 306},  220  (2002), and references
  therein.

\bibitem{Panagiotopoulos99}
A.~Z. Panagiotopoulos and S.~K. Kumar, Phys. Rev. Lett. {\bf 83},  2981
  (1999).

\bibitem{Dickman99}
R. Dickman and G. Stell,  in {\em Simulation and Theory of Electrostatic
  Interactions in Solutions}, edited by L.~R. Pratt and G. Hummer (AIP,
  Woodbury, NY, 1999), p.\ 225.

\bibitem{Brognara02}
A. Brognara, A. Parola, and L. Reatto, Phys. Rev. E {\bf 65},  066113  (2002),
  and references therein.

\bibitem{Fisher72}
M.~E. Fisher, S. k.~Ma, and B.~G. Nickel, Phys. Rev. Lett. {\bf 29},  917
  (1972).

\bibitem{Luijten02}
E. Luijten and H.~W.~J. Bl{\"o}te, Phys. Rev. Lett. {\bf 89},  025703  (2002).

\bibitem{Katzgraber03a}
H.~G. Katzgraber and A.~P. Young, Phys. Rev. B {\bf 67},  134410  (2003).

\bibitem{Katzgraber03b}
H.~G. Katzgraber and A.~P. Young, Phys. Rev. B {\bf 68},  224408  (2003).

\bibitem{Bray84}
A.~J. Bray and M.~A. Moore, J. Phys. C {\bf 17},  L463  (1984).

\bibitem{McMillan84a}
W.~L. McMillan, Phys. Rev. B {\bf 29},  4026  (1984).

\bibitem{McMillan84b}
W.~L. McMillan, Phys. Rev. B {\bf 30},  476  (1984).

\bibitem{Rieger96}
H. Rieger {\it et~al.}, J. Phys. A {\bf 29},  3939  (1996).

\bibitem{Hartmann01}
A.~K. Hartmann and A.~P. Young, Phys. Rev. B {\bf 64},  180404  (2001).

\bibitem{Carter02}
A.~C. Carter, A.~J. Bray, and M.~A. Moore, Phys. Rev. Lett. {\bf 88},  077201
  (2002).

\bibitem{Fisher88}
D.~S. Fisher and D.~A. Huse, Phys. Rev. B {\bf 38},  386  (1988).

\bibitem{Baranovskii79}
S.~D. Baranovski\u{i}, A.~L. Efros, B.~L. Gel'mont, and B.~I. Shklovski\u{i},
  J. Phys. C {\bf 12},  1023  (1979).

\bibitem{Diaz00}
A. D{\'i}az-S{\'a}nchez {\it et~al.}, Phys. Rev. B {\bf 62},  8030  (2000).

\bibitem{Vojta94}
T. Vojta and M. Schreiber, Phys. Rev. Lett. {\bf 73},  2933  (1994).

\bibitem{Grannan94}
E.~R. Grannan and C.~C. Yu, Phys. Rev. Lett. {\bf 73},  2934  (1994).

\bibitem{Mobius04}
A. M{\"o}bius, private communication.

\bibitem{Mobius03}
A. M{\"o}bius and U.~K. R{\"o}{\ss}ler, Short-range type critical behavior in
  spite of long-range interactions: the phase transition of a Coulomb system on
  a lattice, cond-mat/0309001.

\bibitem{Xue88}
W. Xue and P.~A. Lee, Phys. Rev. B {\bf 38},  9093  (1988).

\bibitem{DeLeeuw80}
S.~W. de~Leeuw, J.~W. Perram, and E.~R. Smith, Proc. R. Soc. Lond. A {\bf 373},
   27  (1980).

\bibitem{Bhatt88}
R.~N. Bhatt and A.~P. Young, Physical Review B {\bf 37},  5606  (1988).

\bibitem{Binder81}
K. Binder, Z. Phys. B {\bf 43},  119  (1981).

\bibitem{privCommCiach}
A. Ciach, private communication.

\bibitem{Vollmayr93}
K. Vollmayr, J.~D. Reger, M. Scheucher, and K. Binder, Z. Phys. B {\bf 91},
  113  (1993).

\bibitem{Young_pc}
A.~P. Young, private communications.

\bibitem{Fisher94}
M.~E. Fisher, J. Stat. Phys. {\bf 75},  1  (1994).

\bibitem{Luijten02b}
E. Luijten, M.~E. Fisher, and A.~Z. Panagiotopoulos, Phys. Rev. Lett. {\bf 88},
   185701  (2002).

\bibitem{Binder86}
K. Binder and A.~P. Young, Rev. Mod. Phys. {\bf 58},  801  (1986).

\bibitem{Mobius92}
A. M\"{o}bius, M. Richter, and B. Drittler, Phys. Rev. B {\bf 45},  11568
  (1992).

\bibitem{Efros76}
A.~L. Efros, J. Phys. C: Solid St. Phys. {\bf 9},  2021  (1976).

\end{thebibliography}
\end{document}